\documentclass[12pt]{iopart}

\usepackage{epsfig}
\usepackage{graphicx}
\usepackage{dcolumn}
\usepackage{bm}
\usepackage{amssymb}
\usepackage{color}
\makeatletter

\newcommand{\Rmnum}[1]{\expandafter\@slowromancap\romannumeral #1@}
\makeatother

\begin{document}

\title[Y. Luo $et$  $al$., CeNiAsO]{CeNiAsO: an antiferromagnetic dense Kondo lattice}

\author{Yongkang Luo$^{1}$, Han Han$^{1}$, Hao Tan$^{1}$, Xiao Lin$^{1}$, Yuke Li$^{1}$, Shuai Jiang$^{1}$, Chunmu Feng$^{1}$, Jianhui Dai$^{1}$, Guanghan Cao$^{1,2}$, Zhu'an Xu$^{1,2}$, and Shiyan Li$^{3}$}

\address{$^{1}$ Department of Physics, Zhejiang University, Hangzhou 310027, P. R. China}
\address{$^{2}$ State Key Laboratory of Silicon Materials, Zhejiang University, Hangzhou 310027, P. R. China}
\address{$^{3}$ Department of Physics and Laboratory of Advanced Materials, Fudan University, Shanghai 200433, P. R. China}

\ead{\mailto{zhuan@zju.edu.cn}}
\date{\today}

\begin{abstract}
A cerium containing pnictide, CeNiAsO, crystallized in the
ZrCuSiAs type structure, has been investigated by measuring
transport and magnetic properties, as well as specific heat. We
found that CeNiAsO is an antiferromagnetic dense Kondo lattice
metallic compound with Kondo scale $T_K \sim$ 15 K and shows an
enhanced Sommerfeld coefficient of $\gamma_0 \sim$ 203
mJ/mol$\cdot$K$^{2}$. While no superconductivity can been observed
down to 30 mK, Ce ions exhibit two successive antiferromagnetic
(AFM) transitions. We propose that the magnetic moment of Ce ion
could align in the G type AFM order below the first transition at
$T_{N1}$=9.3 K, and it might be modified into the C type AFM order
below a lower transition at $T_{N2}$=7.3 K. Our results indicate
that the 3$d$-4$f$ interlayer Kondo interactions play an important
role in Ni-based Ce-containing pnictide.

\end{abstract}


\maketitle

\section{Introduction}

The discovery of iron-based arsenide superconductor
LaFeAsO$_{1-y}$F$_{y}$\cite{Hosono-LaF} has triggered enormous
enthusiasm in searching new high transition temperature ($T_c$)
superconductors in Fe or Ni-based
pnictides\cite{ChenXH-SmOF,WangNL-CeOF,WenHH-LaSr,Hosono-LaNi,LiZ-LaNiAsOF}.
$T_c$ has been raised up to above 50 K when La is replaced by
other rare earth elements in the so-called 1111 type family
\cite{ZhaoZX-LnOD,WangC-GdTh}. The parent compound of the 1111
type pnictides, $LnTPn$O ($Ln =$ rare earths, $T =$ transition
metals, $Pn $= As, P), is of the prototype of ZrCuSiAs structure,
where the conducting $TPn$ layer is sandwiched by the insulating
$Ln$O layer. So far superconductivity (SC) can be induced in the
Fe(Ni)As-based pnictides by many different ways of chemical doping
or by applying high pressure.

Among all these $LnTPn$O's, the members $Ln =$ Ce show extremely
interesting electronic properties. CeFeAsO, as a prototype parent
compound of Fe-based superconductor, is an itinerant bad metal. A
spin-density-wave(SDW)-like antiferromagnetic (AFM) transition of
$d$ electrons was observed around $\sim 140$ K, and the AFM
transition of the Ce$^{3+}$ local moments was observed below 4 K
\cite{Dai-CeNeutron}. SC has been induced by electron doping (i.e.
F-for-O doping) when the SDW long range order of the Fe
3$d$-electrons is suppressed\cite{WangNL-CeOF}. On the other hand,
CeFePO is a paramagnetic (PM) heavy-fermion (HF) system which is
near the ferromagnetic (FM) instability\cite{Geibel-CeP,XuZA-CeP}.
By comparison, CeOsPO shows the Ce$^{3+}$ AFM order at 4.5 K while
CeRuPO is a rare example of the FM Kondo lattice with $T_{C} =$ 15
K\cite{Geibel-CeRu/OsPO}. For the case of $T =$ Co, both CeCoPO and
CeCoAsO show the FM correlated Co-$3d$
magnetism\cite{Geibel-CeCoPO,Ohta-LnCoAsO,Geibel-CeCoAsO}.

It is noted that the nickel arsenide LaNiAsO is a low $T_c$
superconductor and $T_c$ can be enhanced upon F-doping
\cite{Hosono-LaNi,LiZ-LaNiAsOF}. Interestingly, the partial
substitution of Fe by Ni in LaFeAsO, which introduces two more
3$d$ electrons by each Ni$^{2+}$ dopant, can induce SC in a very
narrow doping regime \cite{CaoGH-LaFeNi}. Moreover, the normal
state resistivity shows an upturn behavior at low temperatures,
suggesting a possible Kondo-effect induced semiconductor behavior.
However, there are very few reports on the physical properties of
1111 type NiAs-based pnictides other than LaNiAsO. The early
report did not observe SC above 2 K and suggested a suspicious FM
order in CeNiAsO \cite{WangNL-CeNiAsO}, although its structure and
the valence of Ce ion has been studied \cite{Balnchard-CeNiAsO}.

In this paper, we report the systematic investigation on the
physical properties of CeNiAsO. No evidence of SC is observed for
temperature down to 30 mK. While there is no local magnetic moment
on Ni ions, two successive antiferromagnetic transitions related
to Ce ions are observed. We propose that the Ce moments could
align in "G" type AFM order first at $T_{N1}$ = 9.3 K, and then
transform into "C" type at a lower temperature $T_{N2}$ = 7.3 K.
The Sommerfeld coefficient is about 203 mJ/mol$\cdot$K$^{2}$,
indicating that CeNiAsO is an enhanced correlated Kondo lattice
compound. All these results imply a strong hybridization between
4$f$- and 3$d$-electrons and highlight the important role of the
3$d$-4$f$ interlayer Kondo physics in nickel based cerium
containing pnictide.

\section{Experimental}

Poly-crystalline CeNiAsO sample of high purity was synthesized by
solid state reaction. Ce, Ni, As, and CeO$_{2}$ of high
purity($\geq$ 99.95\%) were used as starting materials. Firstly,
CeAs was presynthesized by reacting Ce discs and As powders at
1320 K for 72 h. NiAs was presynthesized by reacting Ni and As
powders at 970 K for 20 h. Secondly, powders of CeAs, CeO$_{2}$,
Ni, and NiAs were weighted according to the stoichiometric ratio,
thoroughly ground, and pressed into a pellet under a pressure of
600 MPa in an Argon filled glove box. The pellet was packed in an
alumina crucible and sealed into an evacuated quartz tube, which
was then slowly heated to 1450 K and kept at that temperature for
40 h. For comparison, polycrystalline sample of LaNiAsO was also
synthesized, in the similar process, where La$_{2}$O$_{3}$ was
used as the oxygen source.

Powder X-ray diffraction (XRD) was performed at room temperature
using a D/Max-rA diffractometer with Cu-K$_{\alpha}$ radiation and
a graphite monochromator. Lattice parameters were derived by
Rietveld refinement using the programme RIETAN 2000
\cite{Rietan-2000}. X-ray Photoelectron Spectrum (XPS) analysis
was carried out by using a VG ESCALAB MARK \Rmnum{2} device with
Mg-K$\alpha$ ($h\nu =$ 1253.6 eV) nonmonochromatized source
operating in Constant Analysis Energy (CAE, 50 eV) mode, and data
were collected in a step of 0.2 eV. Before taking the XPS data,
the CeNiAsO sample was pre-polished by the argon beam in the
vacuum to make a newly polished sample surface. Electrical
resistivity was measured with a standard four-probe method in a
Quantum Design physical property measurement system (PPMS-9, for
$T \geq$ 2 K) and a dilution refrigerator (for $T <$ 2 K), while
Hall coefficient was measured by scanning field from -5 T to 5 T.
Thermopower was measured by a steady-state technique and a pair of
differential type E thermocouples was used to measure the
temperature gradient. Specific heat was measured by heat pulse
relaxation method in PPMS-9. The dc magnetization measurement was
carried out in a Quantum Design magnetic property measurement
system (MPMS-5) in zero-field-cooling (ZFC) and field-cooling (FC)
protocals under a magnetic field $H$ of 1000 Oe.

\section{Results and discussion}

Fig.~1 shows Rietveld refinement of XRD patterns of CeNiAsO. All
the XRD peaks can be well indexed in to the tetragonal
ZrCuSiAs-type structure with P4/$nmm$ (No.129) space group, and no
obvious impurity trace can be found. The quality factor ($\chi^2$)
of this refinement reaches to as low as 0.88, guaranteeing the
goodness of sample quality. The refined structural parameters are
listed in Tab.~1 with a comparison to CeFeAsO, from which we can
find that CeNiAsO has a slightly longer $a$-axis but much shorter
$c$-axis compared to its iron based counterpart. Because arsenic
is drawn closer to the Ni plane, the thickness of NiAs layer,
$d_{NiAs}$, is much smaller than $d_{FeAs}$, while the angle of
As-$T$-As ($T =$ Fe or Ni, see the inset of Fig.~1) is enlarged.

Fig.~2 shows temperature dependence of dc magnetic susceptibility
of CeNiAsO measured under $H =$ 1000 Oe. The data above 150 K
exhibit good Curie-Weiss behavior, and can be well fitted to
$\chi=\chi_0+C/(T-\theta_W)$, with $\chi_0$ being the temperature
independent item, and $\theta_W$ the Weiss temperature. The
fitting reports that $\theta_W$ = -31.8 K. The obtained effective
moment $\mu_{eff} =$ 2.24 $\mu_B$ is slightly smaller than that of
a free Ce$^{3+}$ ion, 2.54 $\mu_{B}$. For $T <$ 100 K, we observed
a change in the slope of the temperature dependence of inverse
susceptibility (right axis of Fig.~2), and the fitting parameters
are $\mu_{eff} =$ 2.16 $\mu_B$, $\theta_W =$ -25.6 K. The change
in the slope should be ascribed to the crystal electric field
(CEF) effect. It has been demonstrated both theoretically and
experimentally that Ni ions do not exhibit magnetic ordering in
the nickel based
pnictides\cite{LiZ-LaNiAsOF,JiangS-LaFeNi,XuG-LaOMAs,McQueen-LaNiPO,Ronning-Ni2X2}.
Therefore, the Curie-Weiss analysis suggests that the observed
effective moments in CeNiAsO should come from Ce ions. When cooled
to low temperature, a peak in susceptibility is observed around
9.3 K, followed by an upturn when further cooled. Since the Weiss
temperature derived from the Curie-Weiss analysis is negative, it
is reasonable to conclude that Ce moments are ordered
antiferromagnetically below 9.3 K. The AFM ordering of Ce ions is
compatible with the linear field dependence of magnetization (data
not shown here), as well as the small magnitude of susceptibility
in comparison with CeFeAsO \cite{XuZA-CeP}.

One possible interpretation to the noticeable deviation of the
obtained effective moment from that of the free Ce$^{3+}$ ion is
the presence of valence state Ce$^{4+}$ which has no 4$f$-electron
and local moment. To investigate the valence of Ce in this
compound, we thus carried out the XPS experiment. The XPS spectrum
of CeNiAsO is shown in Fig.~3. After subtracting the background,
the XPS spectrum can be well fitted by a combination of ten
Gaussian peaks, i.e., Ce(\Rmnum{3}) = $v^0$+$v'$+$u^0$+$u'$, and
Ce(\Rmnum{4}) = $v$+$v''$+$v'''$+$u$+$u''$+$u'''$, with $u$ and
$v$ corresponding to the spin-orbit split 3$d_{3/2}$ and
3$d_{5/2}$ core holes, respectively. The positions of these peaks
are taken from Ref.\cite{Beche-CeXPS}, where XPS spectrum of
CePO$_4$ (Ce$^{3+}$) and CeO$_2$ (Ce$^{4+}$) were presented. We
found that Ce$^{3+}$ is dominant in CeNiAsO, but there is a small
trace of Ce$^{4+}$ which should account for the small peak around
916 eV \cite{CeRhSb}. An early report on the Ce 3$d$ XPS and Ce
$L_3$-edge XANES analysis of CeNiAsO and CeFeAsO by Balnchard et
al suggested that the valence of Ce should be close to $3+$
\cite{Balnchard-CeNiAsO}. In other Ce containing compounds with
strong Kondo interaction, for example, in CePd$_2$Al$_3$
\cite{CePdAl}, it is common that Ce has a valence very close to
3+. Therefore, although the valence effect may play some role in
the reduction of effective magnetic moment of Ce, a more likely
interpretation to this reduction could mainly come from the CEF
effect, which will be discussed hereinafter.

The transport properties of CeNiAsO are presented in Fig.~4. The
resistivity exhibits several prominent features. Firstly, the
magnitude of resistivity at room temperature, i.e., $\rho_{300
K}$, has a small value of 3.9 $\mu\Omega\cdot$m, which is about
two orders of magnitude smaller than that of CeFeAsO (see in
Tab.~1). The good metallicity of CeNiAsO can be also confirmed by
Hall coefficient $R_H$ measurement, as shown in the lower inset.
$R_H$ remains as a nearly temperature independent constant above
150 K, and converges to 0.68$\times$10$^{-4}$cm$^{3}$/C. A rough
estimate from single band model leads to a carrier density of
$\sim$10$^{24}$cm$^{-3}$, which is an upper limit of charge
carrier density in a multi-band system, about 3 orders of
magnitude larger than those of
$Ln$FeAsO\cite{McGuire-CePrNd,Tao-Nernst,Ishida-carrier}.
Secondly, in contrast to LaNiAsO, the resistivity in CeNiAsO shows
a hump around 100 K. This could be ascribed to the f-electron
contribution via Kondo scattering between the localized 4$f$- and
the conduction 3$d$-electrons. We subtract the resistivity of
LaNiAsO from CeNiAsO, and the difference is shown in the upper
inset. A broad maximum at 92 K becomes obvious. Similar broad peak
is also observed in temperature dependent thermopower $S(T)$. Such
a phenomenon is a distinct feature in Kondo lattice system, and
can be explained by the Kondo scattering from different CEF
levels\cite{Amato-S,Zlatic-S}. However, we can not exclude the
possibility of onset of coherent Kondo scattering, provided that
both have a comparable temperature scale. Thirdly, when
temperature decreases down to around 9.3 K, a pronounced drop is
observed, which is reminiscent of the reduction of spin-flip
scattering when Ce moments become AFM ordered. Such a drop in
resistivity was also observed in previous report
\cite{WangNL-CeNiAsO}. Moreover, as further cooled, a small kink
emerges around 7.3 K. We associate this kink with the
transformation of Ce magnetic structure proposed hereinbefore.
Finally, while LaNiAsO is a superconductor with $T_c \sim$ 2.75 K
\cite{LiZ-LaNiAsOF}, no SC in CeNiAsO is observed down to 30 mK,
implying the influence of Kondo coupling on the conduction
carriers. A small residual resistivity of $\rho_0=$0.24
$\mu\Omega\cdot$m is derived, which again ensures the high purity
of the sample. It should be noted that both $R_{H}$ and $S$ are
positive above 150 K, and become negative below 100 K, manifesting
a multi-band nature of the system in which hole-type carrier
dominates at high temperature while electron-type charge carrier
becomes dominant at low temperature. Whether this change of
carrier type is CEF effect related needs more investigations.

The result of specific heat measurement is shown in Fig.~5. For
high temperature, $C(T)$ follows good Dulong-Petit law and
saturates to the classical limit of 4$\times$3$R\sim$100
J/mol$\cdot$K (data not shown here). Considering the contributions
from the phonons and the Schottky anomaly, the specific heat can
be written as:
\begin{eqnarray}
C=\gamma_0 T+\beta T^3+C_{Sch}
\end{eqnarray}
for temperatures below $\Theta_D$/10, where $\Theta_D$ is the
Debye temperature, $\gamma_0$ and $\beta$ are coefficients of the
electron and phonon contributions, while $C_{Sch}$ is the Schottky
anomaly item. We first subtract the specific heat of LaNiAsO from
the total specific heat, based on the assumption that both of them
have the similar phonon contribution, to estimate the contribution
from $4f$ electrons in CeNiAsO, and the difference ($C^{4f}$) is
shown in the inset of Fig.5(a). $C^{4f}/T$ does not go to zero at
higher temperature ($T
>$ 30 K), but shows a broad peak centering at 50 K. This broad
peak should be attributed to the Schottky anomaly caused by the
excitations between different CEF levels. A Schottky anomaly
formula with three Kramers doublets (one doublet ground state and
two excited doublets)\cite{Geibel-CeRu/OsPO, Dai-CeCEF} is used to
fit this broad peak, i.e.,
\begin{eqnarray}
C_{Sch}(T) = \frac{R}{g_0+g_1\exp(-\Delta_1/T)+g_2\exp(-\Delta_2/T)} \nonumber \\
~~~~~~~~~~~~\times\{g_0g_1(\Delta_1/T)^2\exp(-\Delta_1/T)+g_0g_2(\Delta_2/T)^2\exp(-\Delta_2/T) \nonumber \\
~~~~~~~~~~~~+g_1g_2[(\Delta_2-\Delta_1)/T]^2\exp[-(\Delta_2-\Delta_1)/T]\}
\end{eqnarray}
where $g_{i} $= 2 is the degeneracy of the $i$th doublet state,
and $\Delta_i$ is the energy difference between ground state and
the $i$th excited state. The obtained energy differences from the
fitting are $\Delta_1 =$ 10.5 meV ($\sim$ 120 K) and $\Delta_2 =$
32.8 meV ($\sim$ 380 K). This result is consistent with the slope
change in $1/\chi(T)$, as well as the observed pronounced broad
maximum observed in both $\rho_{Ce}-\rho_{La}$ and $S$. Due to the
large $\Delta_2$, it is likely that the reduction of effective Ce
moment is caused by the CEF effect. We then subtract the fitted
Schottky anomaly item from the total specific heat, and the result
shows a good linear dependence in the $(C-C_{Sch})/T-T^2$ plot
(see the inset of Fig.~5(b)) for the temperatures below 30 K. The
extrapolated Sommerfeld coefficient $\gamma_0 \sim$ 203
mJ/mol$\cdot$K$^2$, which is more than 40 times of that of LaNiAsO
(see in Tab.~1), manifesting the correlated effect contributed
from the Ce 4$f$-electrons. Although the phonon contribution and
the CEF effect is hard to be removed completely, our analysis
should provide a good estimation of the Sommerfeld coefficient of
CeNiAsO. The linear fit also produces that $\beta =$ 0.2511
mJ/mol$\cdot$K$^4$ and Debye temperature $\Theta_D =$ 314 K, which
indicates the above analysis is quite self-consistent. For $T <$
10 K, two $\lambda$-shaped peaks are observed on the $C/T$ curve,
implying two successive phase transitions. This can be further
confirmed by the derivative of susceptibility ($Td\chi/dT$), the
derivative of resistivity ($d\rho/dT$), the magnetoresistivity
($\rho_{5T}-\rho_0$), and the Hall coefficient ($R_H$) shown in
Fig.~5(b,c). The excellent agreement among $C/T$, $d\rho/dT$ and
$Td\chi/dT$ leads to a definition of two characteristic
temperatures, i.e., $T_{N1} =$ 9.3 K and $T_{N2} =$ 7.3 K.
Actually, similar analysis has been widely applied in other nickel
containing compounds, e.g., CeNiGe$_3$ \cite{Mun-CeNiGe3} and
$Ln$Ni$_2$B$_2$C \cite{Ribeiro-RNiBC}. Note that the magnetic
entropy gain derived by integrating $C^{4f}/T$ over temperature
reaches 70\% of $R ln2$ at 10 K, and recovers the full doublet
ground state ($R ln2$) at 30 K. All these indicate that both
specific heat peaks are due to Ce-$4f$ electrons related magnetic
transitions, though the ordered moment is partially screened by
the Kondo coupling in the ground state doublet. The Kondo scale is
estimated to be $T_K \sim$ 15 K by the entropy.
A ratio of $T_K/T_{N1} \sim$ 1 can also be obtained by judging
from the specific heat jump $\Delta (C^{4f}-C_{Sch})$ at $T_{N1}$
\cite{Besnus-TK/TN}, thus it is reasonable that the Kondo scale
$T_K \sim$ 15 K. We propose that the transition at 9.3 K is
originated from the Ce-AFM transition based on the magnetization
measurement. We also notice that the similar phenomenon was also
found in other Ce-based Kondo lattice compounds such as
CeCu$_2$(Si$_{1-x}$Ge$_{x}$)$_2$, where the two peaks were
explained by two incipient instabilities in the magnetic structure
of pure CeCu$_2$Ge$_2$ \cite{Trovarelli-CeCuSi,HQYuan}: the first
one at a higher temperature, related to a reorientation of the
moments, and a second one at lower temperature, related to a
lock-in of the propagation vector.

Now we turn to discuss the possible magnetic configurations of
CeNiAsO. First of all, we should emphasize that all the observed
magnetism should arise from the Ce moments. The negative value of
the Weiss temperature ($\theta_W$) indicates that Ce moments are
AFM correlated. The most simplified model involves the intra-layer
and inter-layer magnetic interactions between the nearest neighbor
Ce moments, denoted by $J^{intra} = J_{0} + J^{intra}_{RKKY}$ and
$J^{inter} = J^{inter}_{RKKY}$, respectively. Where, $J_{0}$ is
the superexchange interaction via O$^{2-}$ and As$^{3-}$ anions in
the absence of $d$-$f$ coupling \cite{DaiJH-CeFe}, while
$J^{intra}_{RKKY}$ and $J^{inter}_{RKKY}$ are the
Ruderman-Kittel-Kasuya-Yosida (RKKY) interactions mediated by
conduction electrons via $d$-$f$ coupling. Generally,  $J_{0}$ is
always negative ($ J_{0}<0$) and favors AFM ordered ground
state\cite{DaiJH-CeFe}, while $J^{intra}_{RKKY}$ and
$J^{inter}_{RKKY}$ are functions of Kondo coupling ($J_K$) and
density of state at Fermi energy($N_{E_F}$) \cite{Doniach}.  We
can assume $J^{inter}_{RKKY}\sim J^{intra}_{RKKY}=J_{RKKY}$, since
they have almost the same $k_F$ and the same nearest neighbor
distances \cite{note2}. $J_{RKKY}$ may be either negative
(antiferromagnetic) or positive (ferromagnetic) depends on
$J_KN_{E_F}$, which may account for either a G-type or C-type
magnetic order respectively when cooled down (see the lower inset
of Fig.~1). Actually, the changes in the Hall coefficient and
thermopower at low temperatures indeed imply a temperature-induced
change, either quantitatively or qualitatively, in the either
density or relaxation time of different types of charge carriers.
Thus it is plausible that a sign change occurs in $J_{RKKY}$ when
temperature is lowered down to $7K\sim 9K$, resulting in the
magnetic transition from the G-type to C-type magnetic
configuration within the Ce-AFM phase. This scenario provides a
possible explanation on the two magnetic phase transitions, and it
is consistent with the drop in the magnetic susceptibility
($\chi(T)$) in the G-type state and then an upturn as it enters
C-type state at $T < T_{N2}$. Further experimental studies on the
single crystalline sample of CeNiAsO and neutron diffraction
measurement are necessary to testify this scenario.

Finally, it is interesting to compare CeNiAsO with its neighbors,
CeFeAsO and CeCoAsO. Previous first principles calculation on
La$T$AsO \cite{XuG-LaOMAs} suggested that Fe and Co ions show
SDW-like and FM instabilities at low temperatures respectively,
while Ni ions display nonmagnetic behavior. For Ce$T$AsO, Fe ions
in CeFeAsO indeed show SDW-like AFM instability and Co ions in
CeCoAsO are FM ordered\cite{Ohta-LnCoAsO,Geibel-CeCoAsO},
meanwhile neither local magnetic moment nor long range magnetic
order of Ni ions is observed in CeNiAsO. However, this does not
mean that the Kondo coupling between Ce moment and conduction
carriers in the NiAs layer is weaker than that in the FeAs layer.
The situation could be totally on the contrary. We expect the
Ce-4$f$ level is much closer to the conduction band in CeNiAsO
than those in CeFeAsO and CeCoAsO. Thus an enhanced hybridization
between 4$f$- and 3$d$-electrons is possible in CeNiAsO due to the
smaller $c$-axis lattice constant (see Tab.~1), as evidenced by
the moderately large Sommerfeld coefficient and the absence of SC.

\section{Conclusion}

To summarize, the magnetic properties and specific heat of
poly-crystalline CeNiAsO sample with high purity have been
studied. Although CeNiAsO is much more metallic compared to its
neighbor CeFeAsO, no superconductivity is observed down to 30 mK.
Two successive antiferromagnetic transitions of Ce 4f electrons
are observed. We propose that the Ce moments could align in G-type
AFM at $T_{N1}$=9.3 K first, and then modify into C-type at a
lower temperature $T_{N2}$=7.3 K. Regarding to the large
Sommerfeld coefficient of about 203 mJ/mol$\cdot$K$^{2}$, CeNiAsO
represents a new example of AFM dense Kondo lattice with Kondo
scale $T_K \sim$ 15 K in a wide class of the rare earth pnictides
of the ZrCuSiAs-type \cite{Rottgen}. A strong hybridization
between 4$f$- and 3$d$-electrons is proposed and its influence on
the ground state is discussed.

\section*{Acknowledgments}

The authors would like to thank Qimiao Si, Huiqiu Yuan and Shuang
Jia for helpful discussion. This work is supported by the National
Science Foundation of China (Grant Nos 10634030 and 10931160425),
the Fundamental Research Funds for the Central Universities of
China (Program No. 2010QNA3026), and the National Basic Research
Program of China (Grant Nos. 2007CB925001 and 2010CB923003).

\section*{References and Notes}

\pagebreak[4]

\begin{table}
\tabcolsep 0pt \caption{Comparison among CeFeAsO, CeNiAsO and
LaNiAsO. Atomic positions: $Ln$ (1/4, 1/4, $z_{Ln}$), $T$ (3/4, 1/4,
1/2), As (1/4, 1/4, $z_{As}$), O (3/4, 1/4, 0). Some data of CeFeAsO
and LaNiAsO are taken from
Ref.\cite{XuZA-CeP,McGuire-CePrNd,Hosono-LaNi,Tao-Nernst}.}
\vspace*{-12pt}
\begin{center}
\def\temptablewidth{1.0\columnwidth}
{\rule{\temptablewidth}{1pt}}
\begin{tabular*}{\temptablewidth}{@{\extracolsep{\fill}}cccc}
$LnT$AsO                           & CeFeAsO~   &CeNiAsO~   &LaNiAsO \\
\hline
$a$~({\AA})                         & 4.0002    & 4.0767     & 4.1231\\
$c$~({\AA})                         & 8.6412    & 8.1015     & 8.1885\\
$z_{Ln}$                           & 0.1411    & 0.1465     & 0.1470\\
$z_{As}$                           & 0.6547    & 0.6434     & 0.6368\\
$d_{TAs}$~({\AA})                   & 2.6736    & 2.3235     & 2.2404\\
$\theta_{As-T-As}$~(\textordmasculine)& 112.5   & 120.6      & 123.0\\
$R_{wp}$                           &11.96\%    &  8.11\%    & 7.52\%\\
$R_{p}$                            & 8.16\%    &  5.73\%    & -\\
$\chi^2$                           & 1.12      & 0.88       & 2.02\\
$\rho(300K)$~($\mu\Omega\cdot$m)    & 305.8     & 3.9        &3.1\\
$R_H(300K)$~($10^{-4} cm^3/C$)      &  $<$0     & 0.68      & -1.0\\
$S$(300 K)~($\mu$V/K)               & -7.2     & 2.9        & 0.62\\
$\gamma_0$~(mJ/mol$\cdot$K$^2$)     & $\sim$ 59 & $\sim$ 203 &$\sim$ 5 \\
\end{tabular*}
{\rule{\temptablewidth}{1pt}}
\end{center}
\end{table}

\newpage

\begin{figure}
\includegraphics[width=10cm]{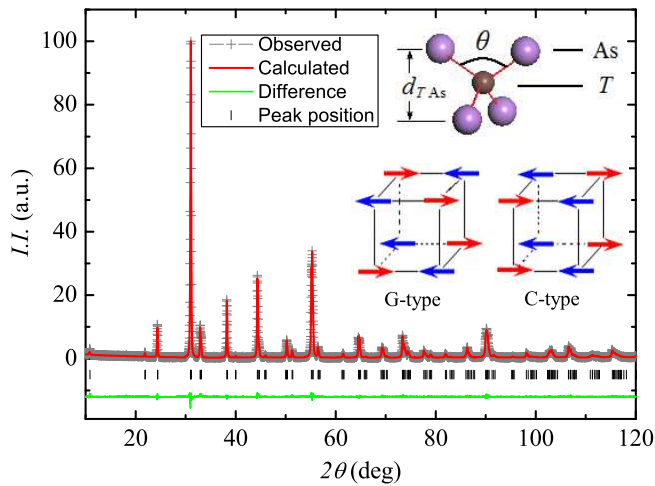}
\caption{(Color Online) Rietveld refinement of CeNiAsO XRD pattern.
Upper inset shows the schematic diagram of $T$As layer. The sketches
of the proposed G type and C type magnetic structures are also shown
in the lower inset.}
\end{figure}

\begin{figure}[htbp]
\includegraphics[width=10cm]{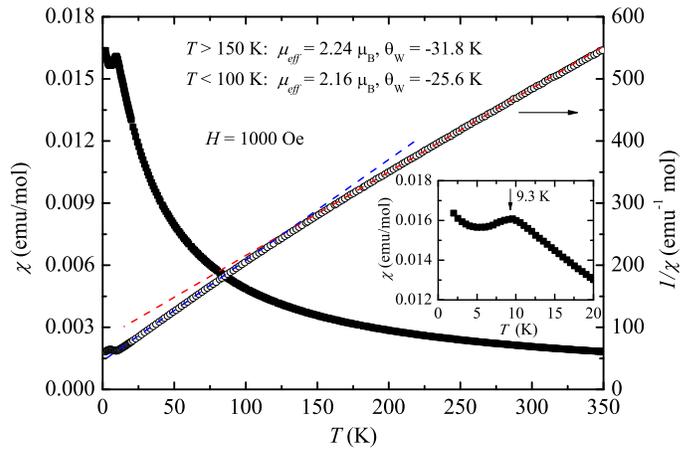}
\caption{(Color Online) Temperature dependence of magnetic
susceptibility of CeNiAsO. The two dashed lines signify Curie-Weiss
fit for $T>$150 K (red) and $T<$100 K (blue), respectively. Inset:
Enlarged plot of $\chi(T)$ at $T <$ 20 K.}
\end{figure}

\begin{figure}[htbp]
\includegraphics[width=10cm]{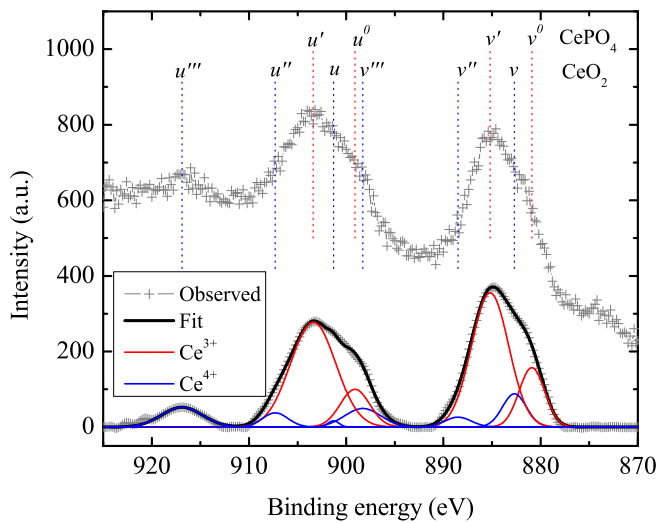}
\caption{(Color Online) Room temperature XPS spectrum.
Ce(\Rmnum{3}) = $v^0$+$v'$+$u^0$+$u'$, and Ce(\Rmnum{4}) =
$v$+$v''$+$v'''$+$u$+$u''$+$u'''$. On the assumption of
intermediate valence of Ce, the XPS spectrum was fitted to a
combination of ten Gaussian functions.}
\end{figure}

\begin{figure}[htbp]
\includegraphics[width=10cm]{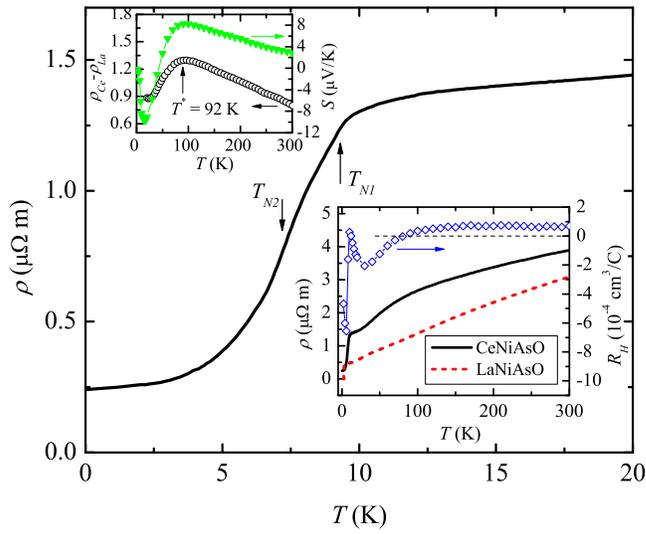}
\caption{(Color Online) Electrical transport properties of
CeNiAsO. Main frame: resistivity at low temperature ($T <$ 20 K),
with two arrows signifying the two magnetic transitions at 9.3 K
and 7.3 K. Lower inset: temperature dependent resistivity of
LaNiAsO and CeNiAsO, and Hall coefficient of CeNiAsO. Upper inset:
The difference of resistivity between CeNiAsO and LaNiAsO,
$\rho_{Ce}-\rho_{La}$, as well as the thermopower $S$.}
\end{figure}

\begin{figure}
\includegraphics[width=10cm]{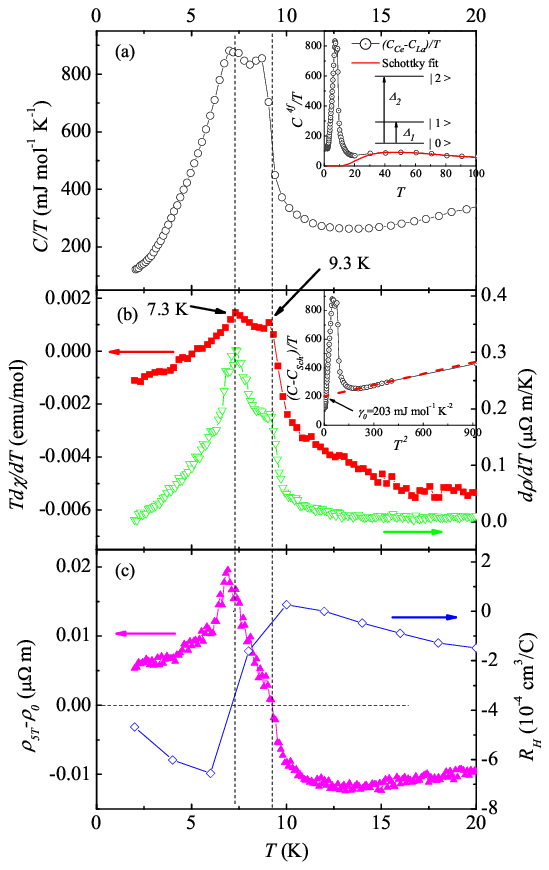}
\caption{(Color Online) Evidences for the two magnetic transitions
below 10 K, from (a), specific heat; (b), derivative susceptibility
($Td\chi/dT$) and derivative resistivity ($d\rho/dT$); (c),
magnetoresistivity ($\rho_{5T}-\rho_0$) and Hall coefficient
($R_H$). Inset of (a) displays the Schottky anomaly fit of
$C^{4f}/T$. Inset of (b) shows $\gamma_0$ derived from extrapolating
$(C-C_{Sch})/T-T^{2}$ plot to zero limit. }
\end{figure}


\begin{thebibliography}{10}
\bibitem{note2}
We argue that $J_{RKKY}$ depends on the Ce-Fe-Ce distance, thus
inter-layer and intra-layer RKKY interactions have almost the same
Ce-Fe-Ce distances.
\end{thebibliography}
\end{document}